\documentclass[aps,prd,onecolumn,nofootinbib]{revtex4}  
\usepackage[paperwidth=8.5in,paperheight=11in,top=1in, bottom=1in, left=1.in, right=1.in]{geometry}
\usepackage{graphicx}  
\usepackage{dcolumn}   

\makeatletter
\let\cat@comma@active\@empty
\makeatother

\usepackage{graphicx}
\usepackage{amsmath}
\usepackage{amsfonts}
\usepackage{epsfig}
\usepackage{slashed}
\usepackage{braket}

\usepackage{comment}
\usepackage{cases}

\usepackage[utf8]{inputenc}
\usepackage{amsmath}
\usepackage{dcolumn}
\usepackage{float}
\usepackage{physics}
\usepackage{appendix}
\usepackage{breqn}
\usepackage{graphicx}
\usepackage[pdftex]{hyperref}
\usepackage[dvipsnames]{xcolor}
\usepackage{comment}
\usepackage{amssymb}
\usepackage[normalem]{ulem}

\newcommand{\be}{\begin{eqnarray}}
\newcommand{\ee}{\end{eqnarray}}

\newcommand{\calO}{\mathcal{O}}

\begin{document}

\title{Simulating the Femtouniverse on a Quantum Computer}

\author{Nouman Butt\footnote{ntbutt@illinois.edu}}
\author{Patrick Draper\footnote{pdraper@illinois.edu}}
\author{Jiayu Shen\footnote{jiayus3@illinois.edu}}
\affiliation{
Illinois Quantum Information Science and Technology Center, Urbana, IL 61801\\
Illinois Center for Advanced Studies of the Universe, Urbana, IL 61801\\
Department of Physics, University of Illinois at Urbana-Champaign, Urbana, IL 61801}

\begin{abstract}
We compute the low-lying spectrum of 4D SU(2) Yang-Mills in a finite volume using  quantum simulations. In contrast to small-volume lattice truncations of the Hilbert space, we employ toroidal  dimensional reduction to the ``femtouniverse" matrix quantum mechanics model.  In this limit the theory is equivalent to the quantum mechanics of three interacting particles moving inside a 3-ball with certain boundary conditions. We use the variational quantum eigensolver and quantum subspace expansion techniques to compute the string tension to glueball mass ratio near the small/large-volume transition point, finding qualitatively good agreement with large volume Euclidean lattice simulations. 
\end{abstract}
\maketitle

\section{Introduction}
Quantum simulations of quantum field theories (QFTs) in the noisy intermediate-scale quantum (NISQ) era~\cite{Preskill2018quantumcomputingin} will be limited by the gap between the complexity of QFTs and the available quantum resources. Currently, it is difficult to build a comparison between known Monte Carlo and quantum simulation results. In particular,  gauge theories are challenging to simulate on quantum devices because of the infinite tower of states associated with each link on the lattice. To control the rapid growth in degrees of freedom requires working with small lattices and putting rather stringent truncations on the subspaces of the Hilbert space associated with bosonic degrees of freedom. Subsequently one must take care to preserve the Gauss law and confront various limitations on what can be computed with small circuits. It is therefore interesting to explore a wide variety of approximations, truncations, encodings, and simulation techniques to make the most of available near-term systems. In the process we may hope to develop new applications and tools that will help lower the threshold for quantum supremacy in simulating high energy physics phenomena, while creating a large set of benchmark computations for calibrating quantum simulations against analytic and classical results.

In this paper we  study hybrid quantum simulations of  an approximation to 4D gauge theory known as the femtouniverse~\cite{Bjorken:1979hv}. Complementary to the small-volume lattice approach, the approximation we use is a matrix quantum mechanics model obtained by the dimensional reduction of the 4D theory on a spatial 3-torus. The study of gauge theory on a torus was pioneered by 't Hooft~\cite{thooft} and the effective matrix quantum mechanics theory obtained by dimensional reduction was systematically developed by L\"uscher~\cite{Luscher:1982ma}, L\"uscher and M\"unster~\cite{LUSCHER1984445}, van Baal~\cite{VANBAAL1986548,van1988gauge,vanBaal1990}, van Baal and Koller~\cite{vBK,van1987qcd}, and others.
For a comprehensive review see~\cite{van2001qcd}. We  focus on the SU(2) model studied in detail by van Baal and Koller~\cite{vBK,van1987qcd}. In our quantum simulations we use the variational quantum eigensolver (VQE)~\cite{peruzzo2014variational,tilly2022variational} and quantum subspace expansion (QSE)~\cite{PhysRevA.95.042308} techniques. These are hybrid classical-quantum  methods that exploit the strengths of both types of computation. With these methods we obtain the low-lying spectrum of the theory, including glueball masses and string tensions, which can be compared with the large-volume continuum limit results known from Euclidean lattice Monte Carlo~\cite{Teper}.

In small volumes, the effective theory from integrating out Kaluza-Klein modes is accurate due to asymptotic freedom. However, it differs markedly from the large-volume limit. At large torus volumes, the spectrum of the complete 4D theory is quite insensitive to the volume. There is a ``large-small volume  transition region'', roughly around $\Lambda L\sim\calO(1)$, where some strong-coupling physics of the large volume theory is reflected with reasonable accuracy in the much simpler effective theory. This is the regime we would like to probe.

Our work is organized as follows. 
In Section~\ref{sec:matrix_model} we begin by reviewing the effective Hamiltonian described in~\cite{vBK} and the symmetries of the effective theory, which allow the decomposition of the Hamiltonian into different superselection sectors. In Section~\ref{sec:qse} we briefly review the QSE algorithm used in the calculation of excited state energies.  Section~\ref{sec:ed} presents our main results. We compute ground and excited state energies using classical exact diagonalization (ED), VQE+QSE simulators, and the IBM-Lima quantum computer. We discuss effects of Hilbert space truncation and different basis choices and various sources of error. We find that the string tension to glueball mass ratio of the continuum large-volume theory is reasonably well captured by VQE+QSE on the real device, working at couplings near the large/small volume transition.

The study of matrix quantum mechanics in quantum simulations is motivated both as a simpler approximation to higher-dimensional nonabelian gauge theories, and by  quantum gravity, where matrix models arise in various contexts. Interesting prior work on this subject includes~\cite{gharibyan2021toward,PRXQuantum.3.010324}. In contrast to these works we use a minimal, gauge-invariant Hilbert space and focus on the explicit model that arises from dimensional reduction on a 3-torus, including the loop-induced effective potential and gauge field topology associated with the parent 4D theory. Dimensional reduction is also a useful approach for truncating abelian theories while preserving some structure; previously we have studied the reduction of the Schwinger model to the quantum mechanical particle on the circle model in quantum simulation~\cite{PhysRevD.105.074505}, where dynamics associated with an 't Hooft anomaly and the $\theta$ term~\cite{Hooft1980} are transparent.

\section{Matrix model for \texorpdfstring{$4D$ $\mathrm{SU}(2)$}{4D SU(2)} gauge theory\label{sec:matrix_model}}
We begin by reviewing the physics of the femtouniverse. In this section, we rely heavily on the pioneering work of van Baal and Koller~\cite{van1987qcd,vBK}. We consider pure SU(2) Yang-Mills theory in 4D on a small spatial 3-torus of length $L$ and work in $A_0=0$ gauge. The Hamiltonian is
\begin{equation}
   H = \int_{[0, L]^3}
   d^3 \mathbf{x} \bigg( \frac{1}{2}g^2 E^{a}_{k}(\mathbf{x}) E^{a}_{k} (\mathbf{x}) + \frac{1}{2g^2} B^{a}_{k}(\mathbf{x})B^{a}_{k}(\mathbf{x})\bigg)
   \label{eq:fullham}
\end{equation}
where $E^{a}_{k}$, $B^{a}_{k}$ are chromo-electric and chromo-magnetic fields, and $g$ is the dimensionless strong coupling. Here $k$ denotes the spatial index and $a$ labels the color index. The gauge connection $A^a_{k}$  is taken to satisfy periodic boundary conditions on the torus.\footnote{Working in sectors of nonzero magnetic flux would be an interesting direction for generalization.}

From lattice analysis we know the theory is gapped near the dynamical scale 
$\Lambda =\mu e^{-8\pi^2/(b g^2(\mu))}$
and becomes exponentially insensitive to $L$ for $\Lambda L \gtrsim 1$. The effective field theory (EFT) analysis below, on the other hand, will be valid for $\Lambda L\lesssim 1$, where $g$ may be thought of as the running coupling at the renormalization group scale $1/L$. Thus we will be particularly interested in the behavior near the small volume-large volume transition.

The proper gauge transformations of the SU(2) theory are periodic SU(2)-valued functions $g(\mathbf{x})$ acting on the connections as 
\be
A_{k}(\mathbf{x}) = g(\mathbf{x}) A_{k}(\mathbf{x}) g^{-1}(\mathbf{x}) -   i g(\mathbf{x}) \partial_{k} g^{-1}(\mathbf{x})
.
\ee 
The classical vacuum manifold, sometimes called the ``vacuum valley'', is the space of flat connections modulo small gauge transformations. The vacua separate into a union of disjoint sectors characterized by different Chern-Simons numbers and related by large gauge transformations. We will work in the sector of fixed vanishing Chern-Simons number, which is acceptable as long as the states of interest have energies below the ``sphaleron'' energy barrier between these sectors. 

We may partially fix the gauge so that the classical vacua are given by spatially constant connections that are aligned in $su(2)$ space. For convenience of illustration we can go to a gauge where the vacua are of the Abelian form 
\be
A_{i} = \frac{C_i}{L}\frac{\sigma_3}{2}.
\label{eq:abeliangauge}
\ee
At this point, it appears that the vacuum valley (in a fixed Chern-Simons sector) is $\mathbb{R}^3$. However, there are still residual gauge transformations, given by 
\begin{align}
g(\mathbf{x}) &= \exp(-4\pi i \frac{\mathbf{x}\cdot\mathbf{n}}{L}\frac{\sigma_3}{2}), \nonumber\\
g &= \sigma_1\;.
\end{align}
Here $n_i$ are integers, and $g=\sigma_1$ is the nontrivial element of the Weyl group.
These gauge transformations lead to the following identifications:
\be
\mathbf{C} \sim \mathbf{C} + 4\pi \mathbf{n} \\
\label{eq:torus}
\mathbf{C} \sim -\mathbf{C}.
\label{eq:orbifold}
\ee
Thus the vacuum valley is the orbifold $T^3/\mathbb{Z}_2$, where the torus periodicity is $4\pi$. 

At the quantum level, the continuous vacuum valley is lifted, but a discrete global symmetry is preserved. This is the $\mathbb{Z}_2$ electric center symmetry, which, in the gauge Eq.~(\ref{eq:abeliangauge}), acts as
\be
h(\mathbf{x}) = \exp(-2\pi i \frac{\mathbf{x}\cdot\mathbf{n}}{L}\frac{\sigma_3}{2})
\ee
where $n_i\in\{0,1\}$. These are not gauge transformations because $h(\mathbf{x})$ and $h(\mathbf{x} + L \mathbf{u}_i)$, where $\mathbf{u}_i$ is the $i$-th spatial unit vector, differ by a nontrivial element of the $\mathbb{Z}_2$ center of $\mathrm{SU}(2)$. They are global symmetries, however, because they preserve the action and the periodic boundary conditions for the gauge field. They act on the classical vacua, which are given by $\mathbf{C} = 2\pi \mathbf{n}$, as
\be
\mathbf{C} \sim \mathbf{C} + 2\pi \mathbf{n}.
\label{eq:abeliangaugecenter}
\ee
Therefore we expect 8 minima of the quantum induced potential on the vacuum valley, lying on the corners of a cube embedded in the 3-torus. We also expect that the symmetry will be unbroken and the true ground state will be similar to a symmetric linear combination of the perturbative ground states around each of these minima. In general the states can be taken to transform in representations
\be
\ket{\psi(A^{h})} = (-1)^{\mathbf{k}\cdot\mathbf{e}}\ket{\psi(A)}
\label{eq:psixform}
\ee
where $\mathbf{e}$ is $\mathbb{Z}_2$-valued electric flux labeling the representations under center~\cite{thooft}.

It is convenient to work with an effective theory that  is partway between the full theory Eq.~(\ref{eq:fullham}) and the vacuum valley theory of the $C_i$ alone. 
We split the gauge field into a linear combination of a spatially constant part and a spatially varying field:
\be
A^{a}_{k}(\mathbf{x},t) = c^{a}_{k}(t) + Q^{a}_{k}(\mathbf{x},t). 
\ee
At small $L$ we can integrate out $Q$ to obtain an effective quantum mechanical theory of the $c^a_k$. This theory includes, in addition to the vacuum valley degrees of freedom, the ``nonabelian'' or ``transverse'' spatially constant modes. Although it seems natural to integrate out Kaluza-Klein modes, it is not obvious that this step is consistent in a gauge theory, where the energy of a mode depends on the vacuum one is perturbing around. The validity of this step hinges on being able to stay relatively close to the origin in field space, with boundaries and suitable boundary conditions at points where the constant field effective theory breaks down. We discuss this further below.

It is also convenient to relax our identification of the vacuum valley with the $\sigma^3$ direction. By a constant gauge transformation we can ``point" the vacuum valley in any direction. We will leave this gauge degree of freedom in the effective theory, then remove it later by averaging over all directions. In this formulation, the vacuum valley corresponds to $c_k$ that are aligned in $su(2)$ space with any common direction. It may be visualized as three aligned (or antialigned) vectors in three dimensions, constrained by the periodicities Eq.~(\ref{eq:orbifold}) to have magnitudes less than $2\pi/L$. The relative angles between the vectors then encode the transverse degrees of freedom.

The effective theory described thus far is the quantum mechanics of three interacting particles in three dimensions. In fact, we can also constrain the particles to move inside a ball, 
\begin{align}
r_i\equiv \sqrt{\sum_a c_i^a c_i^a}\leq \frac{\pi}{L},
\label{eq:ball}
\end{align}
with certain boundary conditions at $\pi/L$:
\begin{align}
    \left.\left(\frac{\partial}{\partial r_i}\right)^{1-e_i}(r_i\psi(c))\right|_{r_i=\frac{\pi}{L}}=0.
    \label{eq:vBKBCs}
\end{align}

The arguments go as follows~\cite{vBK}. The points $r_k\in\{0,\pi\}$ in the vacuum valley are preserved by the center and Weyl transformations. This is clear in the gauge-fixed version of Eqs.~(\ref{eq:orbifold}), (\ref{eq:abeliangaugecenter}) and Eq.~(\ref{eq:ball}) generalizes it to the case where the vacuum valley direction is unfixed. If the wavefunction and the energy is continuous at these points, then either the wavefunction or its (covariant) radial derivative must vanish, depending on the electric flux quantum numbers (cf. Eq.~(\ref{eq:psixform})). 

Continuity, however, is rather subtle. A global center transformation acts as a $2\pi$ translation on the vacuum valley, but makes a nontrivial modification to the spatial dependence of the transverse modes. Since it is a symmetry, these modes must be reordered. The transverse Kaluza-Klein modes that should be kept in a consistent effective theory are completely different from the spatially constant modes valid near the origin, once we have translated by $2\pi$ away from the origin along the vacuum valley. We should think of the points $r_k\in\{0,\pi / L\}$ as boundaries between different patches in which we have different EFTs. More precisely the ``three particles in a ball" description is valid in any patch, but the map to the underlying degrees of freedom of the 4D theory changes from patch to patch. For this reason it is not immediately obvious whether wavefunction continuity (particularly for derivatives) must hold. 

However, Ref.~\cite{vBK} argued that continuity properties are still expected. In brief, at weak coupling, all transverse modes are expected to be close to their ground states at $r_k=\pi/L$, and this property was reflected in the numerical analysis of~\cite{vBK}. At strong coupling, the transverse modes are excited. However, they also mix, and we do not expect level crossing. Continuity of the energy density then leads to the conditions Eq.~(\ref{eq:vBKBCs}).  The interpretation of the wavefunction simply changes discontinuously from patch to patch.

Thus we are led to the quantum mechanics of three particles in the region Eq.~(\ref{eq:ball}) subject to boundary conditions Eq.~(\ref{eq:vBKBCs}). 
The effective Hamiltonian can be written as Eq.~\cite{vBK}
\begin{equation}
    H_{\mathrm{eff}} = - \frac{1}{2L} \bigg( \frac{1}{g^2} + \alpha_1 \bigg)^{-1} \frac{\partial^2}{(\partial c^{a}_{i})^2} + V_{T}(c) + V_{l}(c).
    \label{eq:hamiltonian}
\end{equation}
Here $V_T$ is the ``transverse" part of the effective potential which vanishes along the vacuum valley. $V_l$ is the effective potential along the vacuum valley. These potentials take the form:
\begin{align}
    V_T &= \frac{1}{4}\big(\frac{1}{g^2} +\alpha_2\big) F^{a}_{ij}F^{a}_{ij} +\alpha_3 (F^{a}_{ij}F^{a}_{ij})c^{b}_{k}c^{b}_{k} +\alpha_4 F^{a}_{ij}F^{a}_{ij}c^{b}_{j}c^{b}_{j} +\alpha_5 (\det c)^2 +...\nonumber\\
    V_{l} &= \gamma_1(g) \sum_{i} r^2_i +\gamma_2(g)\sum_{i}r^4_{i} + \gamma_3(g) \sum_{i > j} r^2_i r^2_j+\gamma_4(g) \sum_{i}r^6_{i} + \gamma_5 \sum_{i\neq j}r^4_{i}r^2_{j} + \gamma_6(g) r^2_1 r^2_2 r^2_3 +\gamma_7(g) \sum_{i} r^8_{i}+ ...
\end{align}
where $\dots$ indicates terms at higher orders in the fields and in the loop expansion. We only consider the Hamiltonian  with $ \alpha_1,\alpha_2  \neq 0$ and $\gamma_i(g) = \gamma(0) $ for $i \leq 7$ and ignore higher order terms. The effective theory is truncated at the two derivative order, so it will break down for energies above the Kaluza-Klein scale $\sim 2\pi/L$ (in addition to energies above the sphaleron energy, as discussed above.) 

In order to extract the low-lying spectrum on a quantum simulator we use Hamiltonian truncation and compute the matrix elements classically. The kinetic part can be diagonalized by writing the momentum operator in spherical coordinates for ``particle" $i$ and using a spherical harmonic basis  $Y_{l_i,m_i}(\theta_{i},\phi_{i})$ for each particle.  For the radial part of momentum operator we employ two choices of basis: spherical Bessels $\chi^{(e)}_{n,l}= j_{l}(k^{e}_{n_{i},l_{i}}r)$ with $V(r) =0$ and a harmonic oscillator basis $\chi^{(e)}_{n,l} = r^{l}e^{-\omega r^2/2}M(\frac{1}{2}l+\frac{3}{4}-\epsilon^{(e)}_{n,l}/(2\omega),l+\frac{3}{2},\omega r^2)$ with $V(r) = \frac{1}{2}\omega^2 r^2 $. $M(a,b,z)$ is a confluent hypergeometric function regular at $z=0$~\cite{vBK} and we have set $\omega =1.5$. In either case the radial basis wavefunctions can be chosen to satisfy the Eq.~\eqref{eq:vBKBCs} in each flux-sector labelled by the flux quantum number $e_{i}$. This boundary condition determines the ``momenta'' $k^{e}_{n,l}(\epsilon^{(e)}_{n,l})$ for each particle in terms of the zeros of the chosen basis functions.

In general, VQE performs better the more ``physics" we can inject into the choice of basis. At strong coupling, the wavefunctions tend to spread out more, while for weak coupling they are more concentrated near the origin. For this reason we use the spherical Bessel basis for strong coupling $g \geq 1.2$ and the harmonic oscillator basis for weak coupling $g < 1.2 $. (The spherical bessel basis exactly diagonalizes the tree-level hamiltonian at infinite coupling.) Of course, in the absence of truncation the spectrum is independent of the choice of basis, but with truncations a better choice of basis can minimize the error in eigenvalue computations.

 Since we are using gauge-invariant coordinates on the vacuum valley, we can work in a fully gauge-invariant hilbert space by requiring the angular wavefunctions to be spin-singlets of $\mathrm{SO}(3)$. Hence the gauge invariant wavefunctions have no free $m_i$ quantum numbers. (Therefore the approach here differs from other quantum simulation studies of Yang-Mills type matrix quantum mechanics models where simpler bases are used at the cost of introducing an extended Hilbert space that includes gauge-non-invariant states~\cite{PRXQuantum.3.010324}.)
 
 In sum, the gauge-invariant Rayleigh-Ritz basis for computing the full Hamiltonian matrix are defined as follows:
\begin{equation}
\begin{aligned}
\braket{\{r_i, \theta_i, \phi_i\}_{i=1,2,3} }{l_1l_2l_3n_1n_2n_3;\mathbf{e}} = \sum_{m_1,m_2,m_3} W(l_1l_2l_3m_1m_2m_3)\prod^{3}_{i=1}\chi^{e_i}_{n_i,l_i}(r_i) Y_{l_im_i}(\theta_i,\phi_i) \nonumber\\
    n_{i},l_{i} \in \mathbb{N}, \hspace{0.2cm} |l_1 - l_2| \leq l_3 \leq l_1 + l_2,\hspace{0.2cm} m_i \in \{-l_i, -l_i+1,..., l_i\}
    \label{eq:ritz}
\end{aligned}
\end{equation}
where $W(l_1l_2l_3m_1m_2m_3)$ are the Wigner 3-$j$ symbols. 

Now we must address truncation.  We use eigenvalues of the operator $\frac{\partial^2}{(\partial c^{a}_{i})^2}$ to organize the states in Eq.~\eqref{eq:ritz} in an ascending order. The full Hamiltonian is infinite dimensional so we truncate it to a finite number of states. We can compute the full matrix in Eq.~\eqref{eq:hamiltonian} for finite number of states  by computing angular and radial matrix elements in the Rayleigh-Ritz basis. The effective hamiltonian can be further projected onto the irreducible representations of its  symmetry group. For electric flux sectors $\mathbf{e}=0$ or $\mathbf{e}=(1,1,1)$ the Hamiltonian symmetry group is the full cubic group $O(3,\mathbb{Z})$. The cubic group is a semidirect product $O(3,\mathbb{Z}) = \mathbb{Z}^{3}_2 \rtimes S_3$, where the $\mathbb{Z}_2$ factor corresponds to parity transformations $P_{i}c^{a}_{k} = -\delta_{ik}c^{a}_{k}$ and $S_3$ represents coordinate permutations $\pi$. These symmetries acts on the gauge-invariant states in the following manner:
\begin{align}
    P_{i}\ket{l_1l_2l_3n_1n_2n_3}&=(-1)^{l_i}\ket{l_1l_2l_3n_1n_2n_3} \nonumber\\
    \pi\ket{l_1l_2l_3n_1n_2n_3}&=\ket{l_{\pi(1)}l_{\pi(2)}l_{\pi(3)}n_{\pi(1)}n_{\pi(2)}n_{\pi(3)}}
\end{align}
The cubic group has ten irreps. We focus on the parity even irreps $A^{+}_{1}$(zero flux) and $e^{+}_{1}$ (unit flux, $\mathbf{e}=(0,0,1)$). For $\mathbf{e} \neq 0$, $\mathbf{e} \neq (1,1,1)$ the cubic group is broken to $\mathbb{Z}_2 \rtimes (\mathbb{Z}^{2}_2 \rtimes S_2)$ where $S_2$ permutes the directions with equal electric flux. We construct the Hamiltonians classically in both sectors and numerically diagonalize them for different numbers $M$ of states. Finally, for the case $M=8$ we use VQE and QSE to compute the low-lying spectrum and compare with the classical exact diagonalization.

The ground state in the $A^{+}_1$ sector is used as the reference ground state for the system, identified with the true ground state in the large volume limit. The gap to the first excitation above the ground state in the $A^{+}_{1}$ sector is identified as a glueball mass. The difference in ground state energies in the $e^{+}_1$ flux sector and the $A^{1}_{+} $ sector is identified with the energy of an electric flux string. The spectrum of the Hamiltonian in these irreps suffices to compute observables, like the square root of the string tension to the glueball mass ratio, that can be compared with values from Euclidean lattice simulations.

\section{Quantum subspace expansion for excited states}
\label{sec:qse}

As mentioned above we use VQE to obtain ground states energies. In order to obtain excited state energies, we apply the Quantum Subspace Expansion (QSE) \cite{PhysRevA.95.042308} to the femtouniverse model. QSE is an extension to the VQE algorithm based on quantum measurements of a set of ansatz excitation operators in the optimal ground state estimate $\ket{\psi (\boldsymbol{\theta^*})}$ obtained from VQE. Here $\boldsymbol{\theta} = (\theta_0, \theta_1, ..., \theta_{n_p})$ is the set of parameters in the VQE ansatz and $\boldsymbol{\theta}^*$ denotes the optimal parameters that minimize the energy.

The subspace in QSE is  the Hilbert space spanned by a set of ansatz excitation operators $\{O_i\}_i$ acting on $\ket{\psi (\boldsymbol{\theta}^*)}$, i.e., a set of states $\{O_i \ket{\psi (\boldsymbol{\theta}^*)} \}_i$. To include the ground state itself, we define $O_0 = I$, the identity operator. We denote the dimension of the subspace as $d^{\mathrm{QSE}}$. Then we can evaluate the Hamiltonian on the subspace as
\begin{equation}
    H^{\mathrm{QSE}}_{ij} \equiv \mel{\psi (\boldsymbol{\theta}^*)}{O_i^\dagger H O_j}{\psi (\boldsymbol{\theta}^*)}.
\end{equation}
We further require $O_i$ to be Hermitian operators.
The matrix elements of $H^{\mathrm{QSE}}$ can be measured on a quantum circuit, since the operator $O_i^\dagger H O_j$ can be decomposed as a linear combination of Pauli strings for the original Hilbert space. The basis $\{O_i \ket{\psi (\boldsymbol{\theta})} \}_i$ is not orthonormal, so instead of an ordinary eigenvalue problem for $H^{\mathrm{QSE}}$, we will need to also compute the overlap matrix
\begin{equation}
    S^{\mathrm{QSE}}_{ij} \equiv \mel{\psi (\boldsymbol{\theta}^*)}{O_i^\dagger O_j}{\psi (\boldsymbol{\theta}^*)}
\end{equation}
and solve a generalized eigenvalue problem (GEVP) 
\begin{equation}
    H^{\mathrm{QSE}} \mathbf{v} = \lambda S^{\mathrm{QSE}} \mathbf{v}.
\end{equation}
The eigenvector $\mathbf{v}$ found this way does not directly give the eigenwavefunctions but needs a further transformation explained in Appendix~\ref{sec:details_vqe_qse}. For our purpose, we do not use this transformation and only focus on the energies given by $\lambda$. 
Similar to $H^{\mathrm{QSE}}$, the matrix elements of $S^{\mathrm{QSE}}$ can also be measured on the quantum circuit. When the ansatz excitation operators are chosen effectively, solving  the GEVP yields a set of eigenstates and eigenenergies that are close to the lowest $d^{\mathrm{QSE}}$ eigenstates and eigenenergies of $H$ in the original Hilbert space.

In the example of 3 qubits with the truncation $M = 2^3 = 8$, we choose ${O_i}$ as $O_0 = III$, $O_1 = XII$, $O_2 = IXI$, $O_3 = IIX$ in the Pauli-string notation. This set of ansatz operators becomes approximately accurate when using the spherical Bessel function basis in the large-$g$ limit, since the lowest excitations are approximately the single-qubit-flipped states from the ground state.

Measuring the matrix elements of $H^{\mathrm{QSE}}$ and $S^{\mathrm{QSE}}$ requires evaluations of Pauli strings expectation values at the state $\ket{\psi (\boldsymbol{\theta}^*)}$. Since the total number of Pauli strings in the 3-qubit case is only $64$ and our Hamiltonian is relatively dense, we measure all the $64$ Pauli strings. It is worth noting that the quantum computational resource required at QSE in our case is significantly lower than VQE, because the VQE does quantum measurements at multiple positions in the space of $\boldsymbol{\theta}$, but in QSE, $\boldsymbol{\theta}$ is fixed to $\boldsymbol{\theta}^*$.

The ground state energy obtained from QSE is not necessarily the same as the ground state energy from the prior VQE. We include the results for ground state energy from VQE, and the first excited state energies from the follow-up QSE.

\section{Results from numerical simulation and real quantum hardware\label{sec:ed}}

We implement our numerical simulation on the measurement-based Aer simulator in Qiskit~\cite{Qiskit}, a python framework for quantum computation.  We work in units where the volume is $L=1$ and we scan over the running coupling $g(L)$,  which is equivalent to adjusting the dimensionless combination $\Lambda L$.

We also implement the VQE and QSE computation on the IBM quantum computer Lima. In these computations we use the $M=8$ truncation with 3 qubits and test three values of the coupling, $g \in \{1.8, 2.2, 2.6\}$. We use  measurement error mitigation~\cite{https://doi.org/10.48550/arxiv.2010.08520} to reduce the effect of noise on the real quantum hardware.

The effective Hamiltonian does not exhibit the canonical form of spin or fermionic Hamiltonian. The simplest digital quantum encoding scheme involves expanding the Hamiltonian in a Pauli string (tensor product of Paulis) basis. This naive encoding, for a generic Hamiltonian, leads to number of Pauli strings that is exponential in the number of qubits rather than polynomial.

\begin{figure*}[htb!]
 \centering
 \begin{minipage}{0.48\textwidth}
  \includegraphics[width=\textwidth]{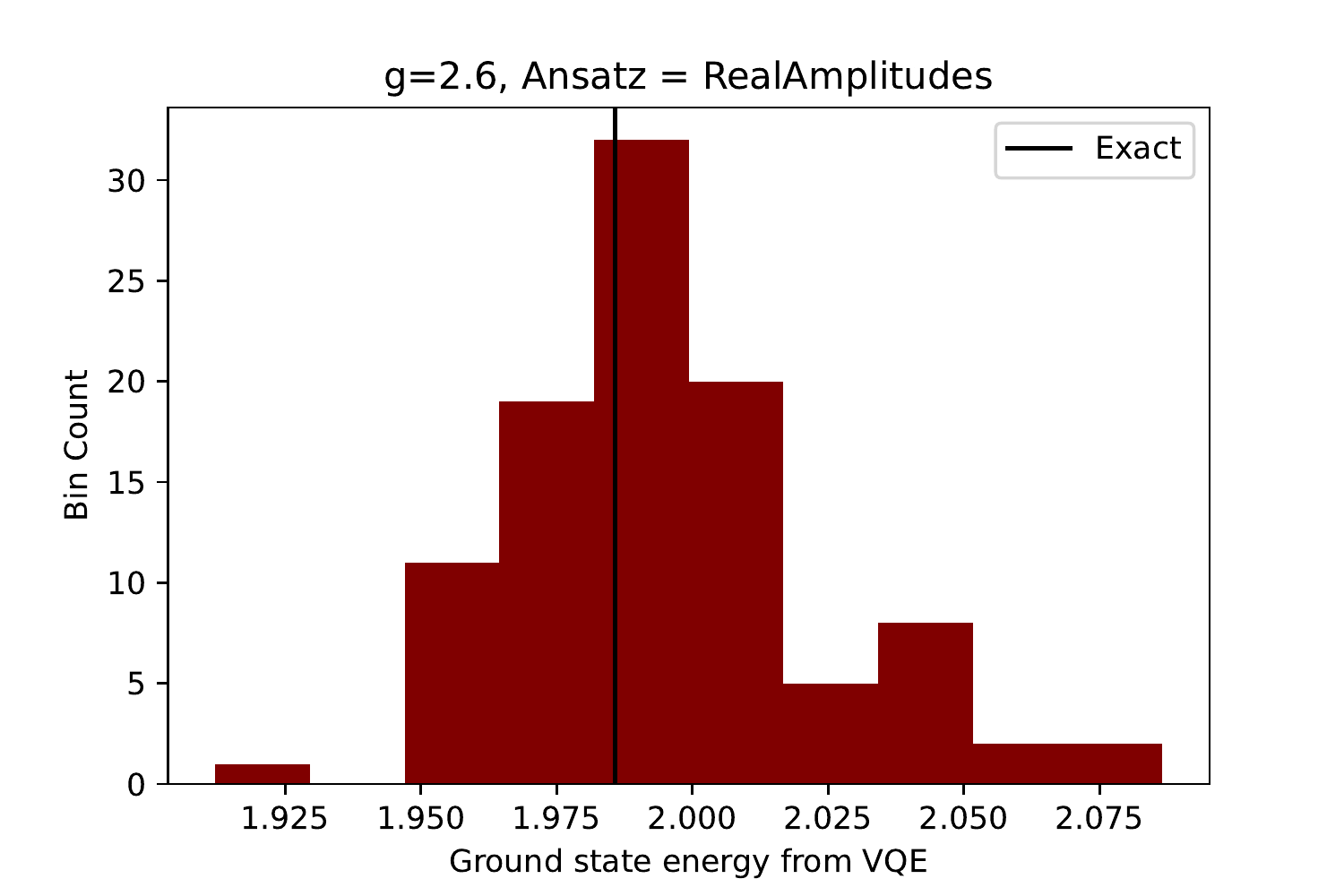}
  \caption{Histogram of 100 noiseless VQE runs vs classical exact diagonalization result at $g=2.6$ for $A^{+}_1$ irrep on Aer simulator with $N_{\mathrm{shots}}=10{,}000$.}
  \label{fig:a1+runs}
 \end{minipage}
 \hfill
 \begin{minipage}{0.48\textwidth}
  \includegraphics[width=\textwidth]{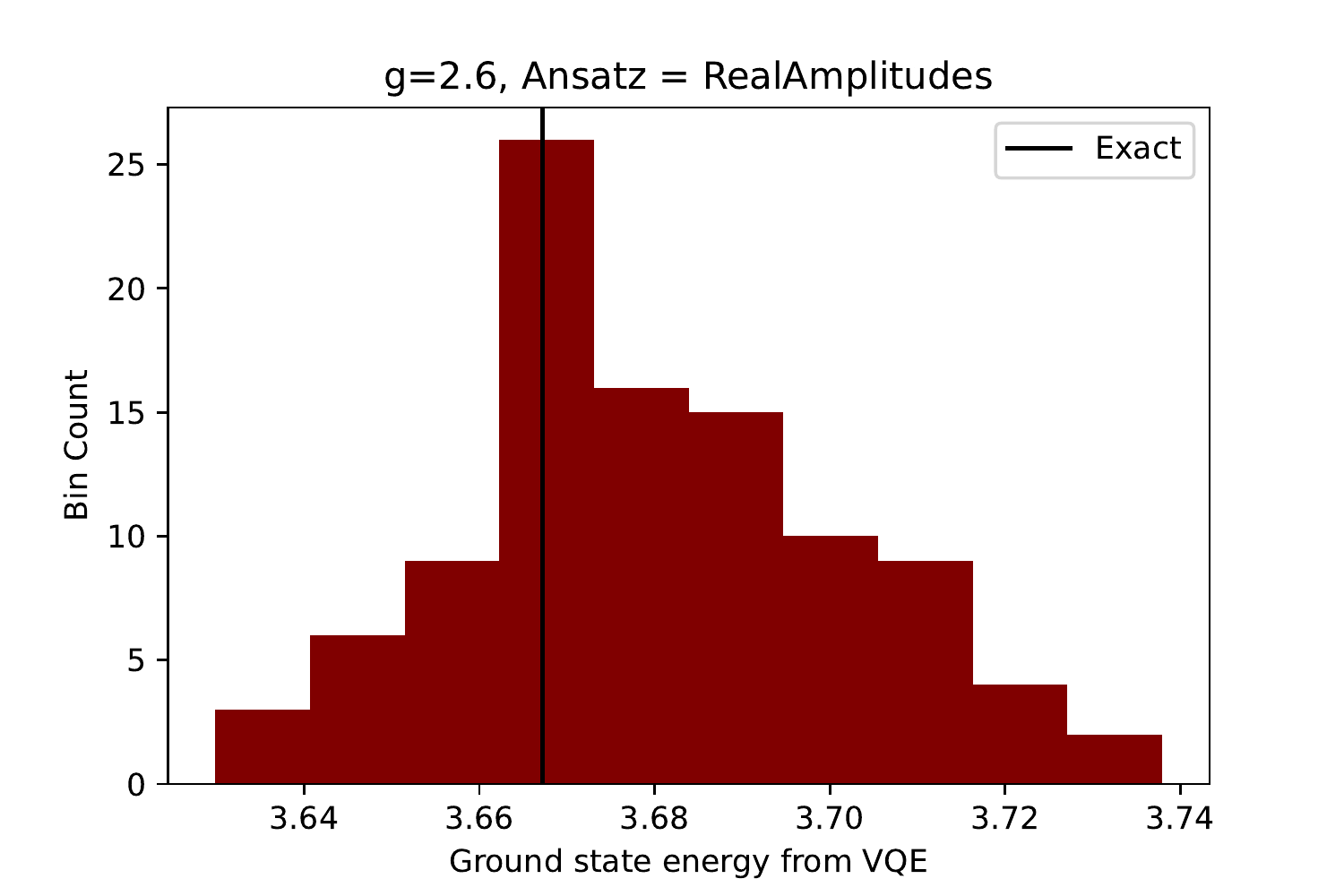}
  \caption{Histogram of 100 noiseless VQE runs vs classical exact diagonalization result at $g=2.6$ for $e^{+}_1$ irrep on Aer simulator with $N_{\mathrm{shots}}=10{,}000$.} 
  \label{fig:e1+runs}
 \end{minipage}
\end{figure*}

We begin by performing multiple runs of noiseless simulated VQE with $N_{\mathrm{shots}} = 10{,}000$ quantum measurements for a given coupling, starting from fully random initial point for every run. This is a test of whether the ansatz is able to reliably find the correct result. In Fig.~\ref{fig:a1+runs} and Fig.~\ref{fig:e1+runs} we show histograms of $A^{+}_1$ and $e^{+}_1$ values obtained from 100 VQE runs with $g=2.6$. The results are clustered around the exact diagonalization value to within a few percent. The fact that VQE can occasionally return results below the truth value is a consequence of shot noise; in our simulations we use 10000 shots, so few-\% errors of this type are expected.

The fidelity between the initial ansatz state and the true ground state, the flatness of energy landscape around local minima~\cite{mcclean2018barren}, and limitations arising from the classical optimizer are the dominant factors which affect the shape of these distributions. In particular the flatness of energy landscape around a local minima can lead to a VQE result which is far away from the rest of the values, even in noiseless simulation.

\begin{figure*}[htb!]
 \centering
 \begin{minipage}{0.48\textwidth}
  \includegraphics[width=\textwidth]{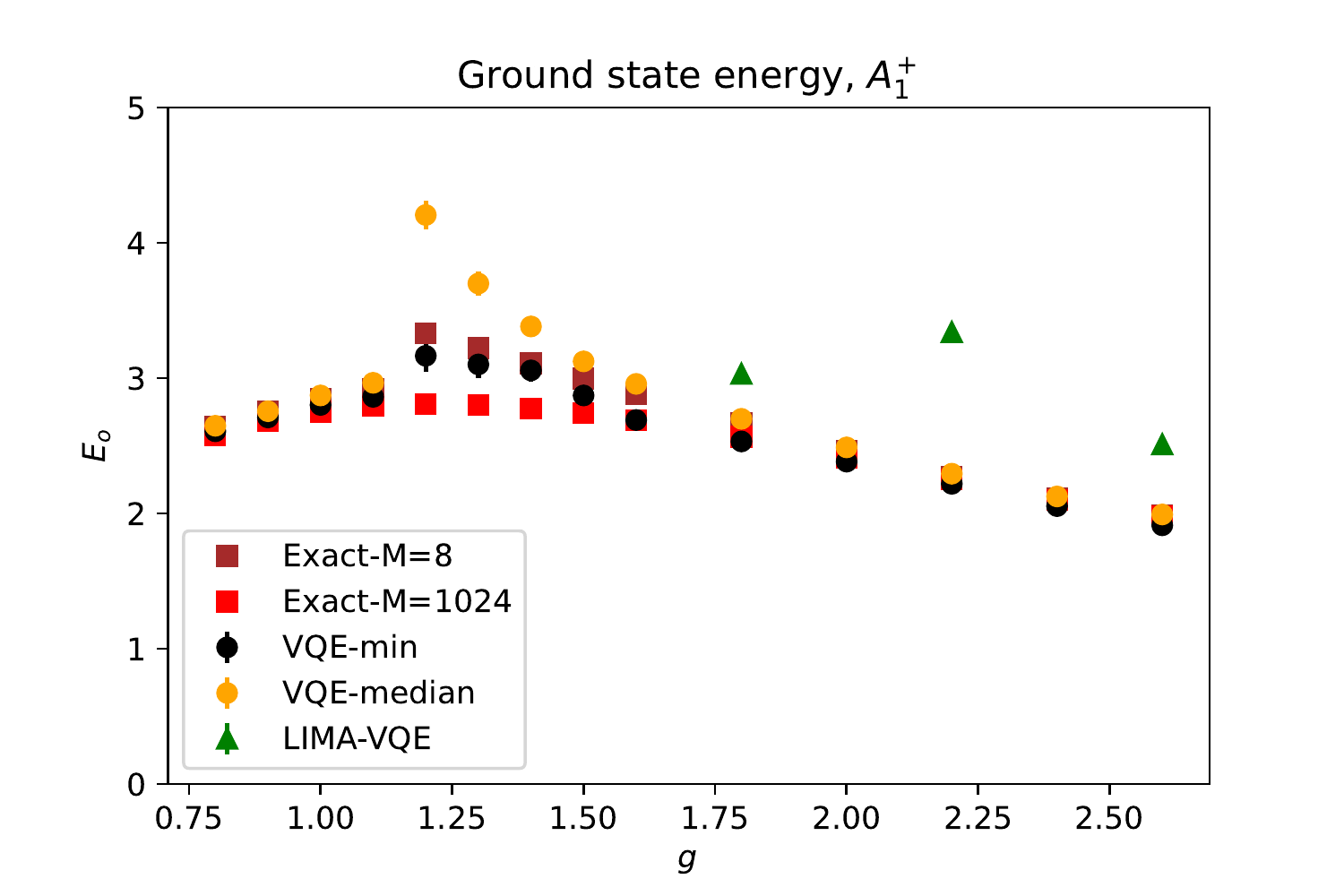}
  \caption{VQE  vs classical exact diagonalization result  for $A^{+}_1$ irrep.}
  \label{fig:a1+}
 \end{minipage}
 \hfill
 \begin{minipage}{0.48\textwidth}
  \includegraphics[width=\textwidth]{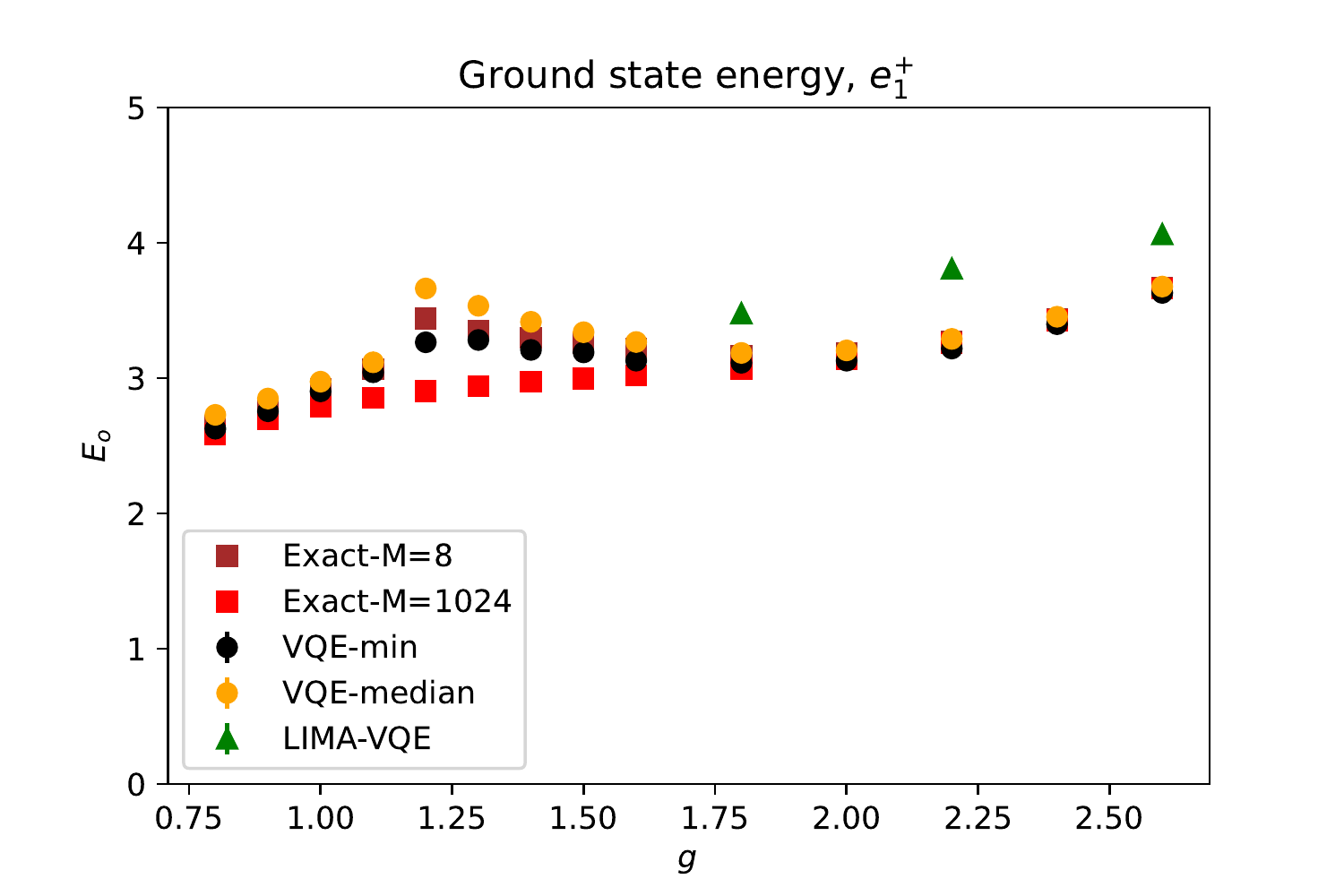}
  \caption{VQE  vs classical exact diagonalization result  for $e^{+}_1$ irrep.} 
  \label{fig:e1+}
 \end{minipage}
\end{figure*}

In Fig.~\ref{fig:a1+} and Fig.~\ref{fig:e1+} we show the noiseless VQE simulation results for a range of $g$ values and compare to the ED values. We also show IBM-Lima results in Fig.~\ref{fig:a1+} and Fig.~\ref{fig:e1+} with $M=8$ states(3-qubits) for $g =1.8,2.2,2.6$. We used the measurement error mitigation~\cite{https://doi.org/10.48550/arxiv.2010.08520} to reduce the effect of noise on the real quantum hardware. We present both the VQE minimum and median\footnote{We order the energies first. When $n_{\mathrm{runs}}$ of VQE is even, the $(n_{\mathrm{runs}} / 2)$-th run (slightly skewed toward the lower side) is chosen as the median, so that the median is always a single sample, not an average of 2 samples. When $n_{\mathrm{runs}}$ of VQE is odd, the usual median is taken.} over 100 runs to emphasize the impact of performing multiple runs.

For most values of $g$, we observe excellent agreement for the ground state results obtained with high-truncation ED  (1024 states), low truncation ED (8 states), and the noiseless VQE simulations with both the minimum and the median taken over runs. However, a clear discontinuity is apparent at $g=1.2$, and for $1.2\lesssim g\lesssim 1.5$, the spread in results is pronounced. This discontinuity arises from the change in basis employed as we go from the weak to the strong coupling regime. The discontinuity is invisible for $M=1024$, indicating that with a large truncation cutoff the results are essentially basis-independent, but the $M=8$ truncation shows a discontinuity. In this case we also see that taking the minimum over VQE runs generally gets closer to the exact value than the VQE median. The disagreement between median and minimum is a consequence of significant numbers of runs in which the VQE algorithm became stuck in spurious local minima, unable to escape due to the flatness of the energy landscape around the local minima. In such cases the minimum is the most reasonable estimate of the ground state energy. However, the mininum can undershoot the exact result due to shot noise, so in some circumstances the median may be preferable.

The IBM-Lima results for $g=1.8,2.2,2.6$ are generally around $\sim 20\%$ above the noiseless simulations and exact results, apart from the result in the $A_1^+$ sector for $g=2.2$ which is significantly higher. The $g=2.2$ experiment was performed on the same quantum computer but on a different day from the $g = 1.8$ and $g = 2.6$ experiments. The larger discrepancy from noiseless simulations exhibited by the $g=2.2$ experiment may reflect a day-to-day variation in the quality of the quantum computer in the lab and calibration of the apparatus therein.

\begin{figure*}[htb!]
 \centering
 \begin{minipage}{0.48\textwidth}
  \includegraphics[width=\textwidth]{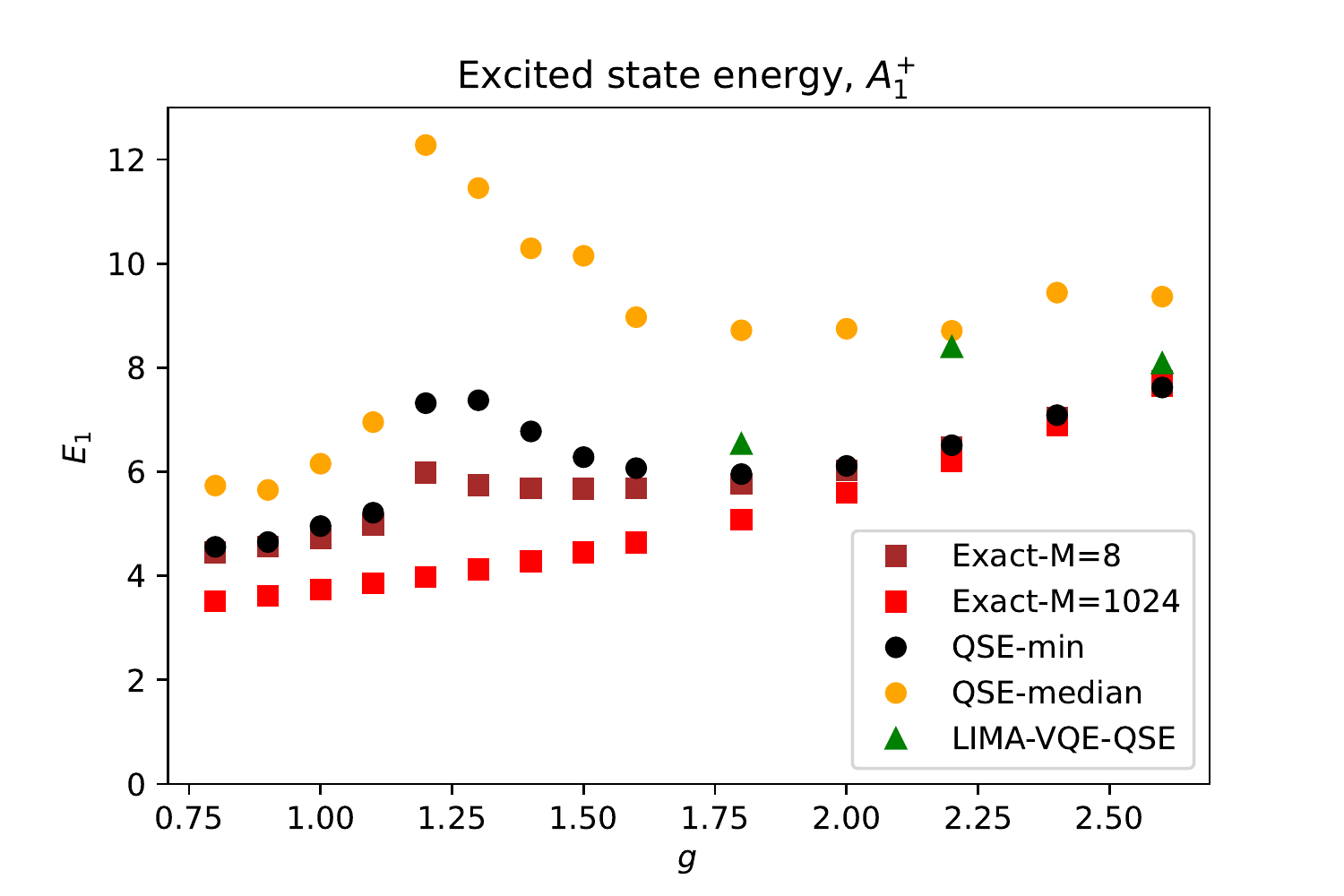} \caption{QSE  vs classical exact diagonalization result  for $A^{+}_1$ irrep excited state.}
  \label{fig:a1+qse}
 \end{minipage}
  \hfill
  \begin{minipage}{0.48\textwidth}
  \includegraphics[width=\textwidth]{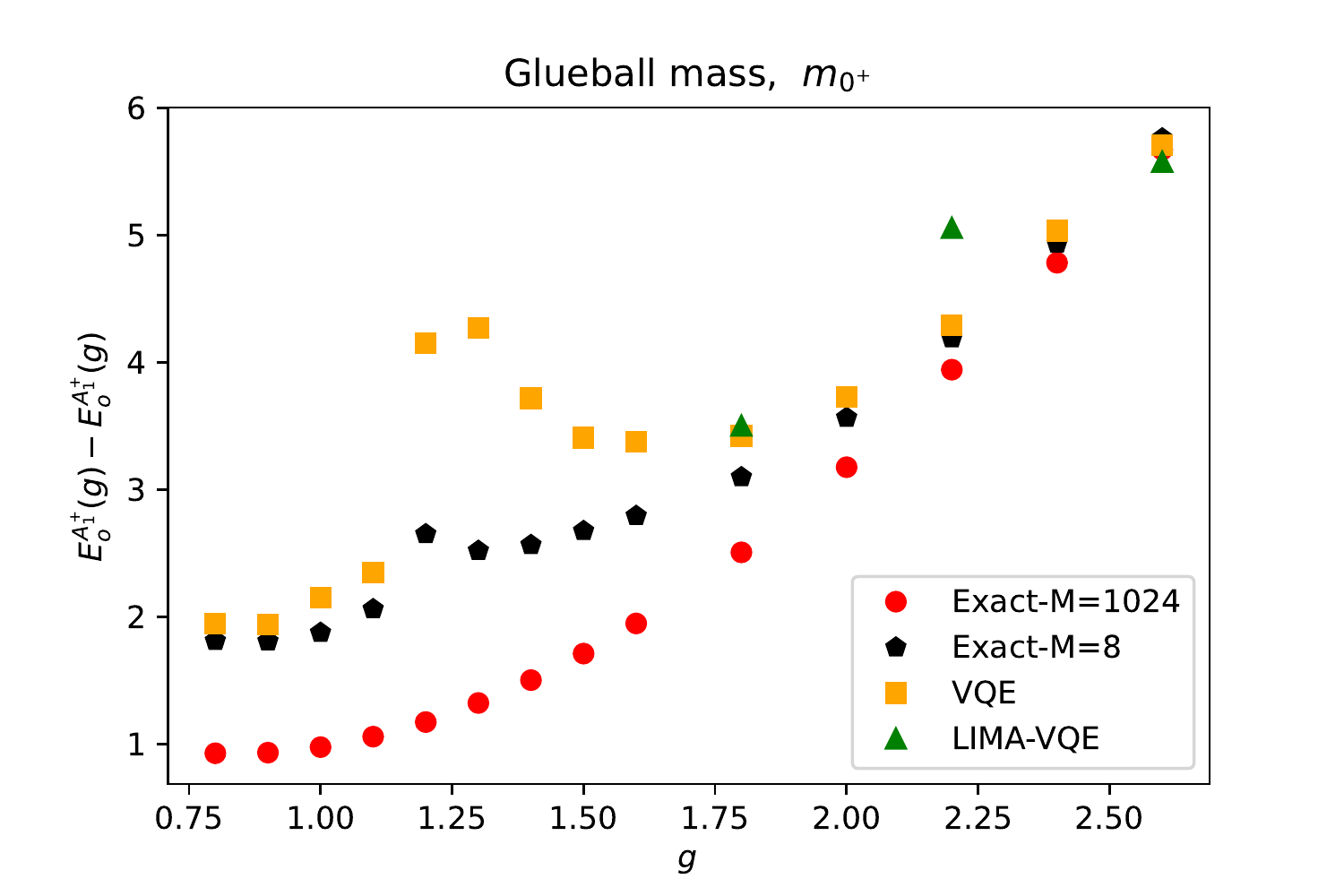}
  \caption{Lightest glueball mass in the $A_1^+$ sector.} 
  \label{fig:glue}
 \end{minipage}
\end{figure*}

Fig.~\ref{fig:a1+qse} shows the first excited state energy in the $A_1^{+}$ sector, comparing the QSE+VQE result  to exact diagonalization. Fig.~\ref{fig:glue}
shows the glueball mass $m_{0^{+}}$, which is the gap between this state and the ground state $m_{0^{+}} = E_1(A_1^{+}) - E_0(A_1^{+})$. The effect of the truncation to $M=8$ states substantially overestimates the glueball mass in the small $g$ region, relative to the $M=1024$ truncation. This situation improves for larger values of $g$.
 We repeat the same procedure of multiple runs using optimal parameters from VQE runs including IBM-Lima runs for $g=1.8,2.2,2.6$. The minimum of all QSE runs is a fairly good estimate of the excited state energy, whereas median values from QSE show large deviations from the exact values. For this reason we show only the minimum results in Fig.~\ref{fig:glue}. These results converge well to the low-truncation exact diagonalization values at both large and small $g$, and to the high-truncation values at large $g$. In addition, the $m_{0^+}$ results on real hardware agree well with the exact results, even in the worst case at $g=2.2$ ($\sim 30\%$), where the upward bias partially cancels in the energy difference.

\begin{figure*}[htb!]
 \centering
 \begin{minipage}{0.48\textwidth}
  \includegraphics[width=\textwidth]{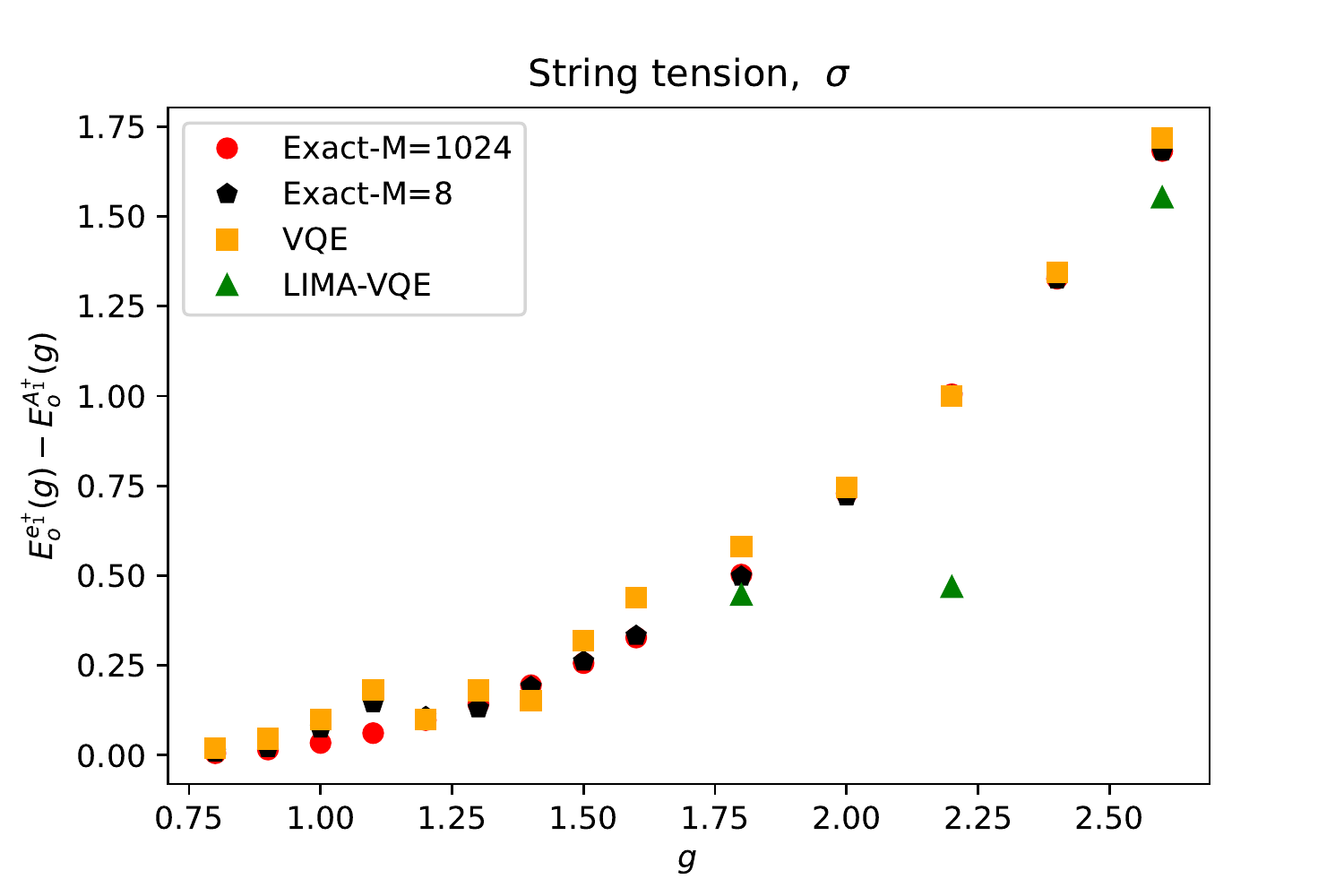} \caption{String tension, $g \in [0.8,2.6]$.}
  \label{fig:string-tension}
 \end{minipage}
  \hfill
   \begin{minipage}{0.48\textwidth}
  \includegraphics[width=\textwidth]{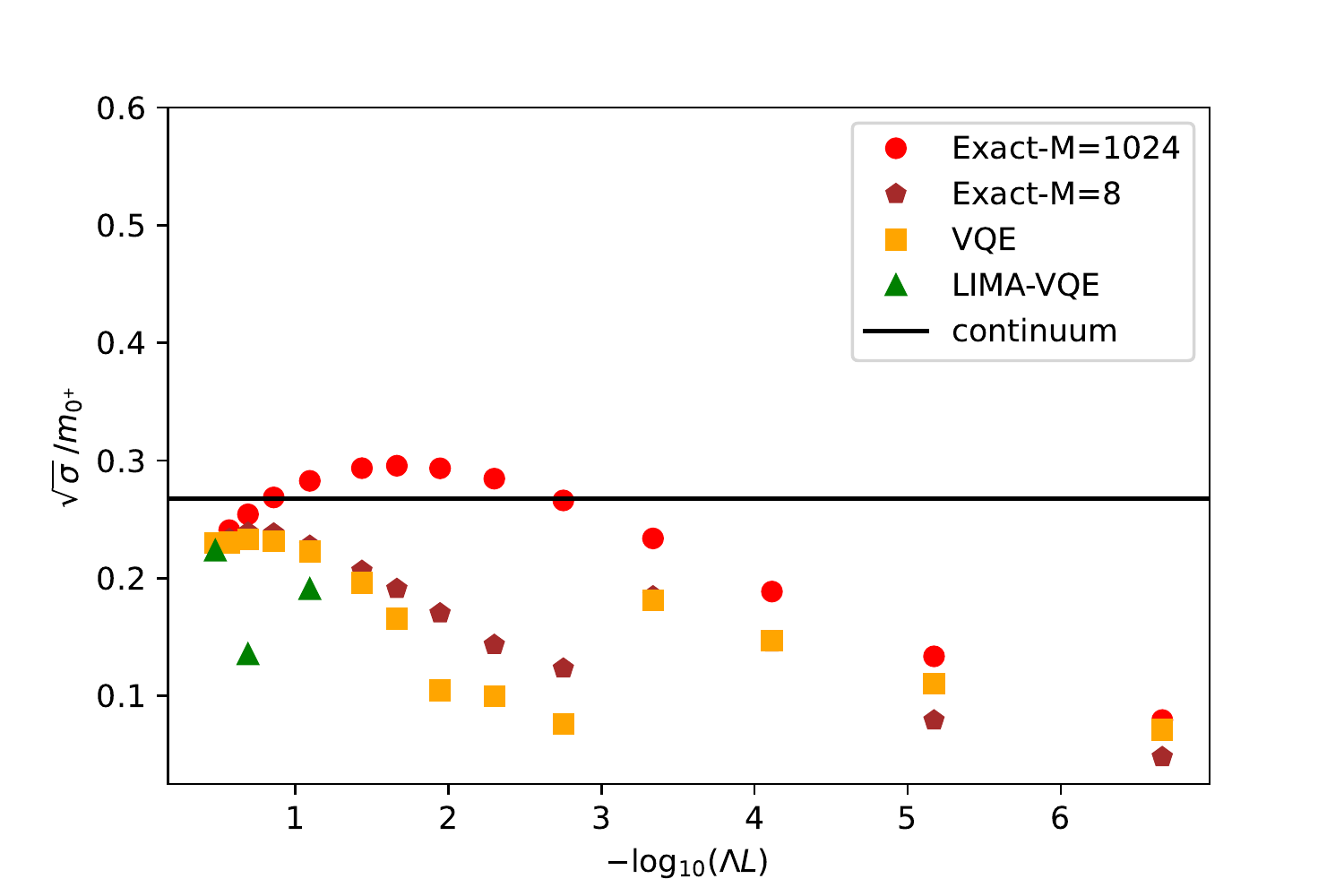}
  \caption{Square root of string tension to glueball mass ratio vs $\log(\Lambda L)$.} 
  \label{fig:glue-string}
 \end{minipage}
\end{figure*}

 In Fig.~\ref{fig:string-tension} we show the string tension $\sigma$ given by the energy gap between the ground states of the $e_1^+$ and $A_1^+$ sectors, $\sigma L = E_0(e_1^+) - E_0(A_1^+)$ (where computationally we work in units with $L = 1$.) The  tension is highly suppressed for small couplings, reflecting the exponential cost of tunneling through the quantum-induced barriers separating center-conjugate vacua. The suppression disappears at larger couplings as wavefunctions are more readily able to penetrate the barrier, signalling the onset of large volume physics.

 We emphasize that the largest values of $g$ we consider coincide with the breakdown of the dimensionally reduced effective quantum mechanics as a good description of the low-lying states of the full 4D theory. Two neglected physical effects come into play around the same time: Kaluza-Klein masses become comparable to the lightest glueball mass and string tension, as does the sphaleron energy barrier separating sectors of varying Chern-Simons number~\cite{van2001qcd}. These are large-volume effects that cannot be captured by the matrix model effective theory. 

\begin{figure*}[htb!]
 \centering
 \begin{minipage}{0.48\textwidth}
  \includegraphics[width=\textwidth]{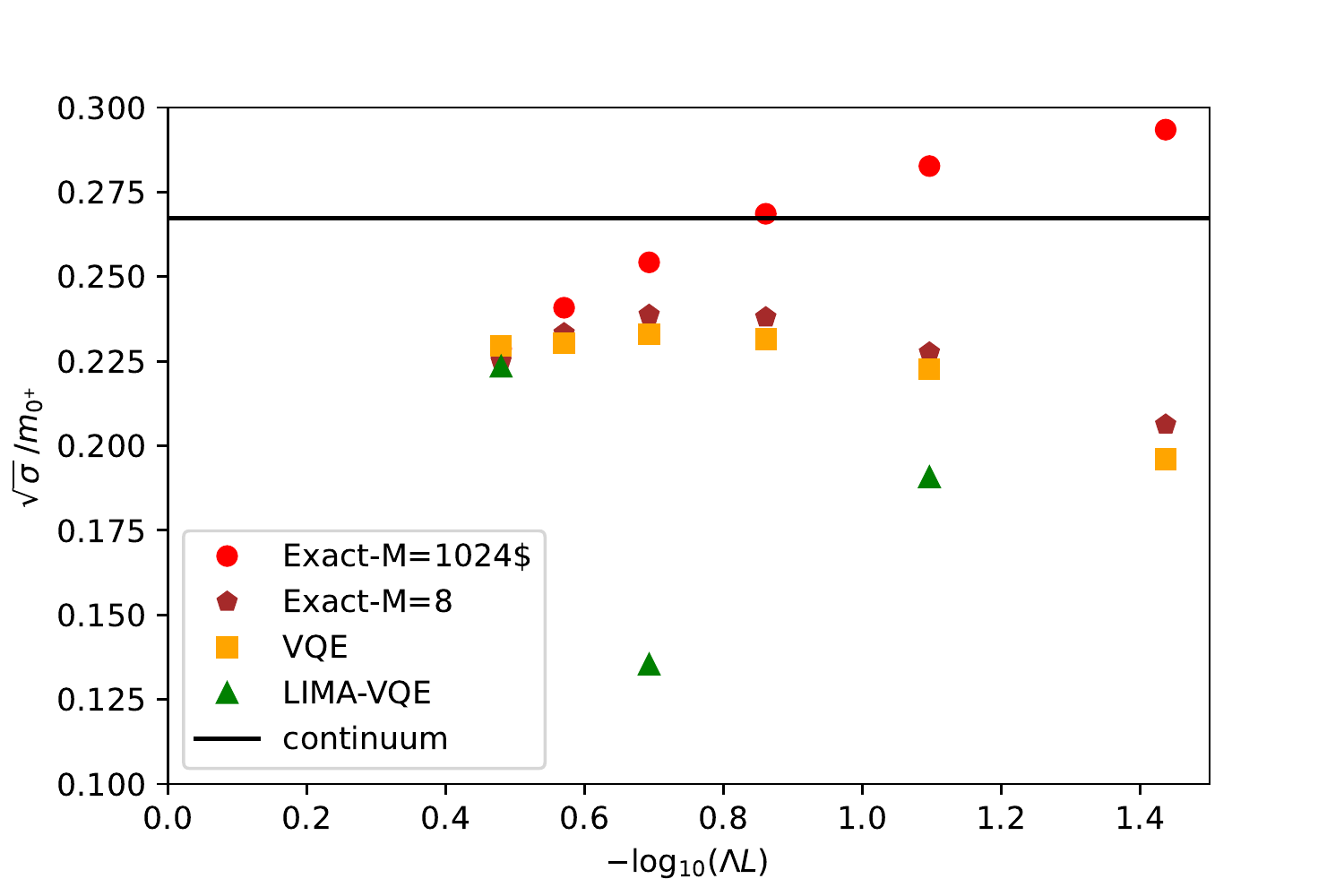} \caption{Square root of string tension to glueball mass ratio.}
  \label{fig:zoom}
 \end{minipage}
 \end{figure*}

Nonetheless, as we approach $\Lambda L\sim 1$ from above, some of the large-volume physics is quantitatively captured by the matrix model. 
In Fig.~\ref{fig:glue-string} we plot the square root of the string tension-to-glueball mass ratio $\sqrt{\sigma}/{m_{0^{+}}}$ against $-\log_{10}(\Lambda L)$, using the 2-loop $\overline{MS}$ estimate for $\Lambda$, $\Lambda L =  e^{-\frac{1}{2\beta_{0}g^2}}(\beta_{0}g^2)^{\frac{-\beta_{1}}{2\beta^2_{0}}}$ , where $\beta_{0} =\frac{N}{16\pi^2}\frac{11}{3}$ and $\beta_1 = (\frac{N}{16\pi^2})^2 \frac{34}{3}$ with $N=2$. We have also added the continuum, large-volume extrapolation from lattice results~\cite{Teper} as a solid horizontal line.  We see that the simulations converge to within about $15\%$ of the continuum, large-volume value for this observable in the largest volumes (left-hand side of the plot), and agreement is found using IBM-Lima for the largest value of $g$. In  Fig.~\ref{fig:zoom} we zoom in on the large $g$ region of this plot.

\section{Conclusions and outlook}

In the near-term, quantum simulations of the type performed here will be limited by noise. We find, for example, that IBM Lima does not give stable results with 4 qubits for our Hamiltonian. This is mainly because of the increased circuit depth and thus the accumulated error from noise. However, there are several directions for improvement that could help push the reach of VQE simulations future, including device-specific choices of gates, optimizing the complexity of the ansatz, quantum error correction~\cite{PhysRevA.32.3266}, and theoretical improvements in the algorithms, such as utilizing commuting families of Pauli strings~\cite{Gokhale2019MinimizingSP} to reduce the number of measurements. Recent developments of Koopman operator learning techniques for quantum optimization~\cite{luo2022koopman} can accelerate VQE when using gradient-based optimizers,  saving quantum resources for other use.

Compared to the lattice approaches, the femtouniverse, as an example of effective theory of the low-momentum modes, represents a different step toward ultimate target of quantum simulations for high energy physics. Using a relatively small number of qubits on a quantum computer, we can already simulate some 4D physics and obtain qualitatively reasonable results in agreement with 4D lattice simulations on classical computers. The dimensional reduction approach supplies a different type of regulator from the lattice regulator, and is relevant to other models of interest in high energy physics including models of quantum gravity~\cite{bfss}. The manifest gauge invariance in the quantum basis also helps reduce the number of dimension of the Hilbert space and thus the number of qubits.

In this paper, using VQE and QSE, we find that the currently available noisy quantum computers can already produce the glueball mass and string tension for the 4D SU(2) Yang-Mills theory qualitatively in agreement with other approaches. Our approach only requires a very small number of qubits, but the Hamiltonian matrix has a relatively large number of elements, and thus may require a relatively deep quantum circuit and big number of Pauli strings.
The encoding of the basis states into the qubit computational basis is not fine tuned, and Hamiltonian matrix elements are all kept without specific ordering or approximation. In addition, with more delicate choice of the representation of the femtouniverse on the quantum computer, it will be interesting to study if the depth of the ansatz can be reduced without loss of essential physics.

The $\mathrm{SU}(2)$ Yang-Mills matrix quantum mechanics model studied here can be extended in a few different directions. Adding higher-momentum modes and their interaction with the zero-momentum modes can improve the approximation to the complete 4D Yang-Mills theory, while adding fermions can connect to more physical models of interest. Future study on how the complexity and performance scales with $N$ for the $\mathrm{SU}(N)$ Yang-Mills theory will likewise inform the scalability and generalizability of the approach.

\subsection*{Acknowledgments}
We thank Aida El-Khadra and Di Luo for helpful discussions. This work was supported in part by the U.S. Department of Energy, Office of Science, Office of High Energy Physics under award number
DE-SC0015655 and by its QuantISED program under an award for the Fermilab Theory
Consortium “Intersections of QIS and Theoretical Particle Physics.” 
We acknowledge the use of IBM Quantum services for this work. The views expressed are those of the authors, and do not reflect the official policy or position of IBM or the IBM Quantum team. In this paper we used \textit{ibmq\_lima}, which is one of the IBM Quantum Falcon Processors.

\appendix

\section{ED results}

Table~\ref{table:energies} contains exact diagonalization results for ground state energies of the irreps $A_1^{+}$, $e^{+}_1 $ and the first excited energy for $A^{+}_1$ denoted by $(A^{+}_1)^{'}$.

\begin{table}[!ht]
\begin{tabular}
{ c @{\hskip 0.4in} c c c @{\hskip 0.4in} c c c @{\hskip 0.4in} c c c } 
\hline\hline
 & & $A_1^{+}$ & & & $(A^{+}_1)^{'}$ & & & $e^{+}_1$ & \\
 $g$ & 1024 & 16 & 8  & 1024 & 16 & 8  & 1024 & 16 & 8 \\
\hline
0.8 & 2.580 & 2.617 & 2.638  & 3.540 & 4.183 & 4.452 &  2.585 & 2.633 & 2.645\\
0.9 & 2.682 & 2.732 & 2.752 & 3.651 & 4.303 & 4.562&  2.697 & 2.758 & 2.772\\
1.0 & 2.752 & 2.821 & 2.844 & 3.763 & 4.486 & 4.722&  2.786 & 2.865 & 2.919\\
1.1 & 2.792 & 2.885 & 2.921 & 3.878 & 4.735 & 4.985 & 2.854 & 2.962 & 3.066\\
1.2 & 2.806 & 2.966 & 3.332 & 3.999 & 4.986 & 5.984 & 2.903 & 3.151  & 3.439\\
1.3  & 2.799 & 2.923 & 3.222 & 4.134 & 4.844 & 5.743 & 2.941 & 3.107 & 3.352\\
1.4 & 2.776 & 2.868 & 3.109 & 4.285 & 4.792 & 5.676 & 2.971 & 3.078 & 3.300\\
1.5 & 2.739 & 2.803 & 2.995& 4.454 & 4.804 & 5.672 & 2.995 & 3.062 & 3.256\\
1.6 & 2.690 & 2.731 & 2.883 & 4.641 & 4.872 & 5.679& 3.018 & 3.058 & 3.216\\
1.8 & 2.565 & 2.578 & 2.668 & 5.074 & 5.164 & 5.768& 3.068 & 3.081 & 3.165\\
2.0 & 2.412 & 2.417 & 2.462 & 5.590 & 5.622 & 6.028 & 3.141 & 3.145 & 3.182\\
2.2 & 2.251 & 2.253 & 2.273& 6.194 & 6.206  &6.463 & 3.256 & 3.258 & 3.273\\
2.4 & 2.101 & 2.102 & 2.110& 6.884 & 6.889 & 7.045& 3.428 & 3.429 & 3.435\\
2.6 & 1.980 & 1.981 & 1.985 & 7.652 & 7.654 & 7.746& 3.663 & 3.664 & 3.667\\
\hline\hline
\end{tabular}
\caption{Ground state and the first excited state energies of the irrep $A^{+}_1$, and the ground state energies of the irrep $e_1^+$ from exact diagonalization with truncations $M = 1024, 16, 8$. As described in the text, we change the basis used for the truncation at $g=1.2$.}
\label{table:energies}
\end{table}

\section{Details in VQE and QSE\label{sec:details_vqe_qse}}

The ansatz wavefunction for the ground state of the Hamiltonian $H$ can be constructed on a quantum circuit as $\ket{\psi (\boldsymbol{\theta})} = U(\boldsymbol{\theta}) \ket{0}$, where $U(\boldsymbol{\theta})$ is a unitary operator consisting of quantum gates parametrized by $\boldsymbol{\theta} = (\theta_0, \theta_1, ..., \theta_{n_p})$, and $\ket{0}$ is the default starting state on the quantum circuit.
The target energy function to be minimized is
\begin{equation}
    E (\boldsymbol{\theta}) \equiv \mel{\psi (\boldsymbol{\theta})}{H}{\psi (\boldsymbol{\theta})}
    .
\end{equation}
$\boldsymbol{\theta}^* \equiv \mathrm{argmin}_{\boldsymbol{\theta}} E (\boldsymbol{\theta})$ gives an approximate ground state of $H$, and the energy evaluated at $\boldsymbol{\theta}^*$  gives the approximate ground state energy.
The Hamiltonian is encoded into the quantum circuit by a linear decomposition in terms of Pauli strings. We also refer to the review~\cite{tilly2022variational} for more details of VQE.

In VQE computations we use the RealAmplitudes ansatz with linear entanglement and $\mathrm{reps}=2$. There are $9$ parameters in the ansatz. RealAmplitudes only gives a real-valued wavefunction, and the eigenstates of the Hamiltonian we consider can always be constrained to be real-valued, so this choice of ansatz is reasonable for our task with a relatively small number of parameters. We use the optimizer COBYLA~\citep{powell1994direct} with $\mathrm{maxiter} = 100$ (maximum number of iterations), $\mathrm{rhobeg} = 0.5$ (reasonable initial changes to the variables), $\mathrm{tol} = 0.001$ (final accuracy in the optimization). The number of shots for quantum measurement is $10{,}000$ so that the statistical error associated with measurement is small. We also use the measurement error mitigation with $10{,}000$ shots when constructing the calibration matrix.

When implementing QSE, we make measurements of all $4^3 = 64$ Pauli strings for the $3$-qubit system, based on the same ansatz as VQE at $\boldsymbol{\theta}^*$. We again use $10{,}000$ shots for each Pauli string evaluation and also the measurement error mitigation.

We denote the  nonorthogonal basis by $\{\ket{\phi_i}\}$ ($i = 0, 1, ..., d^{\mathrm{QSE}} - 1$ where $d^{\mathrm{QSE}} = 4$ in our case).

The transformation between $\{\ket{\phi_i}\}$ and an orthonormal basis $\{\ket{\psi_i}\}$ is
\begin{equation}
    \ket{\psi_i} = P^*_{ij} \ket{\phi_j}
    .
\end{equation}
Then the following operator identity holds
\begin{equation}
    P S P^\dagger = I
    ,
\end{equation}
where
\begin{equation}
    S_{ij} = \braket{\phi_i}{\phi_j}
\end{equation}
is the overlap matrix. In the QSE basis, $S = S^{\mathrm{QSE}}$ in the main text.
An operator $O$ when acting on the nonorthonormal basis transforms as
\begin{equation}
    O' = S P^\dagger O P S
    .
\end{equation}
With the Hamiltonian matrix in the nonorthogonal basis
\begin{equation}
    H^{\mathrm{QSE}}_{ij} = \mel{\phi_i}{H}{\phi_j}
\end{equation}
we need to solve the GEVP
\begin{equation}
    H^{\mathrm{QSE}} \mathbf{v} = \lambda S \mathbf{v}.
\end{equation}
We can transform the eigenvector $\mathbf{v}$ back by
\begin{equation}
    \mathbf{u} = P S \mathbf{v}
    ,
\end{equation}
where $\mathbf{u}$ is the wavefunction of the eigenstate in the original orthonormal basis $\{\ket{\psi_i}\}$.

\renewcommand\refname{References}
\bibliographystyle{unsrt}
\bibliography{main}

\end{document}